\documentclass[11pt]{article}

\input epsf.sty

\def\lromn#1{\uppercase\expandafter{\romannumeral#1}}

\begin{document}

\begin{center}
\begin{Large}
\textbf{
Photon and neutrino-pair emission from 
circulating quantum  ions  
}
\end{Large}

\vspace{2cm}
M.~Yoshimura and N. Sasao$^{\dagger}$

\vspace{0.5cm}
Center of Quantum Universe, Faculty of
Science, Okayama University \\
Tsushima-naka 3-1-1 Kita-ku Okayama
700-8530 Japan

\vspace{0.2cm}
$^{\dagger}$
Research Core for Extreme Quantum World,
Okayama University \\
Tsushima-naka 3-1-1 Kita-ku Okayama
700-8530 Japan \\

\end{center}

\vspace{3cm}
\begin{center}
\begin{Large}
{\bf ABSTRACT}
\end{Large}
\end{center}

The recent proposal of photon and neutrino pair beam
is extensively investigated.
Production rates, both differential and total, 
of single photon, two-photon and  neutrino-pair
emitted from quantum ions in circular motion
are calculated for any velocity of ion.
This part is an extension  of our previous results
at highest energies to lower energies of circulating ions,
and helps much to identify the new process
at a low energy ion ring.
We clarify how
to utilize   the circulating ion for a new source
of coherent neutrino beam despite of much stronger background photons.
Once one verifies that the coherence is maintained in
the initial phases of time evolution after laser irradiation,
large background photon emission rates are not
an obstacle  against utilizing the extracted neutrino pair beam.

\vspace{2cm}

PACS numbers
\hspace{0.5cm} 
13.15.+g, 
14.60.Pq, 

Keywords
\hspace{0.5cm} 
CP-even neutrino  beam, 
heavy ion synchrotron, 
quantum coherence

\newpage

\section
{\bf Introduction}

Photon and neutrino pair emission from circulating ions 
are of great interest and one may describe the process
following the theory of  synchrotron radiation \cite{schwinger} and similar extension
to the neutrino case.
This is legitimate,
when ions are circulated in the ground state.
In these cases produced neutrinos and photons
remain in the low energy region, typically in the  keV
range, hence one can essentially ignore
the neutrino pair emission due to extremely low
production rates.
Their emission spectrum and rates are however completely
different and are much more interesting, when ions are circulated
in a quantum mixture of excited and ground  states, as
demonstrated in our recent paper \cite{pair beam},
which presented results in the high energy limit of ions.
This new mechanism of photon and neutrino pair production
is hoped to provide a new tool of neutrino and photon beam
extracted from  ion ring.

In the present work we shall deepen the understanding 
of the production mechanism of photon and neutrino pair
from quantum ion ring,
and calculate basic quantities of 
production rates  at any velocity .of circular motion.
This is expected to help verify experimentally 
important features of this new process.

The rest of this work is organized as follows.
In Section 2 we recapitulate basic features of
particle emission from quantum ion beam, and
derive the fundamental differential rate for
single photon emission at any ion velocity.
In Section 3 outputs of analytical and numerical results
are presented for the photon
energy spectrum near the forward direction
and the angular distribution, paying a special attention to
how these are changed with ion circulation velocity
or the boost factor.
Section 4 deals with the coherence decay,
which might endanger emission of photons and neutrino pairs.
The problem is important to secure 
a stable  way in which the emission process occurs.
Section 5 discusses the neutrino pair emission  
and how  circulating heavy ions may
become a strong source of the neutrino-pair beam,
regardless of even stronger backgrounds of photon emission.
In Appendix A we discuss the two-photon emission which
may become the source of de-coherence along with
the single photon emission.
In Appendix B we give
some relevant mathematical
items related to the subject in the present work.

Throughout this work we use the natural unit of $\hbar = c = 1$.

\vspace{0.5cm} 
\section
{\bf Photon emission from quantum ion beam: basic features}

We shall extend derivation of
fundamental formulas of photon emission rate of
\cite{pair beam}
to include cases of low velocity (small $\beta = v/c$) regions.

The quantum coherent state of a single ion (in the Schroedinger picture)
is defined by a superposition of two states,  $|e\rangle$ and $|g\rangle$,
\begin{eqnarray}
&&
| c(t) \rangle = \cos \theta_c e^{-i \epsilon_g t/\gamma} |g \rangle 
+ \sin \theta_c e^{-i \epsilon_e t/\gamma} e^{i\varphi_c} | e\rangle
\,.
\end{eqnarray}
We assume $|e\rangle$ to be a metastable excited state, while $|g\rangle$ is the ground state of the ion.
This state is not an energy eigen-state, but may be realized after
laser irradiation.
Without a loss of generality we may take $\varphi_c = 0$, which we shall do in the following.
The boost factor $\gamma = 1/\sqrt{1-\beta^2}$, with $\beta$ the velocity divided by the
light velocity.

We assume that a single ion in this quantum state is circulating in the ring of radius $\rho$
such that its orbit is
\begin{eqnarray}
&&
\vec{r}_I(t) =  \rho (\sin  \frac{\beta t}{\rho}\,,  \cos  \frac{\beta t}{\rho} -1 \,, 0)
\,,
\end{eqnarray}
in a coordinate system of $\vec{r}_I(0) = 0$.
($\beta =v/c$ here and below is meant to be the velocity of circulating ions, with the natural unit of
the light velocity $ c =1$.)
We consider coherent emission of many photons of definite momentum, 
\begin{eqnarray}
&&
\vec{k} = \omega (\cos \psi \cos \theta, \cos \psi \sin \theta, \sin \psi)
\,,
\end{eqnarray}
and some helicity from a circulating single ion.

We  consider the photon emission by electric dipole (E1) interaction.
Extension to the magnetic dipole (M1) transitions is straightforward.
The probability amplitude of E1 photon is given using an expectation value of
$|c(t) \rangle$:
\begin{eqnarray}
&&
{\cal M}(t) = \int_{0}^t dt' \langle c(t') | \frac{e \vec{p}} {m_e} | c(t') \rangle
\cdot \vec{A}(t'; \vec{k}, h) 
\\ &&
=  - i \frac{\epsilon_{eg} } {\sqrt{2 \omega V}}
\int_{0}^t dt' \langle c(t') | \vec{d}  | c(t') \rangle
\cdot \vec{e}_h
e^{i\omega t' - i \vec{k}\cdot \vec{r}_I(t')}
\,.
\end{eqnarray}
$V$ is the quantization volume of emitted electromagnetic field.
This is a coherent sum along the orbit trajectory of a single ion.
The production rate is defined by time derivative of the probability,
hence it is given by
\begin{eqnarray}
&&
\partial_t | {\cal M}(t)|^2 = 2\Re \left( \langle c(t) |  \frac{e \vec{p}} {m_e}  | c(t) \rangle
\cdot \vec{A}(t; \vec{k}, h)
{\cal M}(t)^*
\right)
\,.
\end{eqnarray}

It is important from the fundamental physics points to use the $p\cdot A/m$ gauge instead of
$d\cdot E$ gauge often used in atomic physics calculation.
These two gauges give approximately identical results when
photons are nearly on the mass shell, namely $\omega \sim \epsilon_{eg}$,
but they may give completely different answers when photons are
far off the mass shell, namely $\omega \gg \epsilon_{eg}$ in our problem.
We typically deal with cases of $\omega/\epsilon_{eg} = O(\gamma)$
which may be very large.
Differences of  rates in dependences on the boost factor $\gamma$
are large in the two gauges.

Consider a situation in which
 all emitted photons of definite momentum $\vec{k}$
are collected by some detector.
We sum over all available ions of the number $I 2\pi \rho/Q$
where $I$ is the DC current of heavy ion of charge $Q$.
The total emission rate from all ion sources is 
\begin{eqnarray}
&&
d \Gamma = \frac{V d^3 k}{(2\pi)^3} \frac{2\pi \rho I}{Q}\gamma
\sum_h 
 2\Re \left( \langle c(t) | \frac{e \vec{p}} {m_e} | c(t) \rangle
 \cdot \vec{A}(t; \vec{k}, h)
{\cal M}(t)^*
\right)
\,,
\\ &&
 2V \Re \left( \langle c(t) |  \frac{e \vec{p}} {m_e} | c(t) \rangle
\cdot \vec{A}(t; \vec{k}, h)  {\cal M}(t)^*
\right)
\nonumber
\\ &&
=
\frac{\epsilon_{eg}^2}{\omega} (\sin\theta_c \cos \theta_c)^2
\sum_{pol} e_h^i (e_h^j)^* (d_{eg})_i (d_{eg})_j \cos  \tilde{\Phi}(0)
\int_{0}^t dt'  \cos \tilde{\Phi} (t')
\,,
\\ &&
\tilde{\Phi} (t) = (  \omega - \frac{\epsilon_{eg}}{\gamma})t - \rho \omega \cos \psi
\left( \sin (\theta + \frac{\beta t}{\rho} ) - \sin \theta
\right)
\,.
\label {phase 1}
\end{eqnarray}
The need to insert the boost factor $\gamma$
is explained in \cite{pair beam}.

Note that the emission rate from the excited state
$|e\rangle$ does not have this type of time integral,
since the factor $\epsilon_{eg}/\gamma$ is missing
in eq.(\ref{phase 1}) in that case.
It leads to the usual synchrotron radiation
\cite{schwinger}
from the state $|e\rangle$, and
the emitted photons have much lower energies than
of order 1 keV (and much smaller rates in the neutrino pair emission).
On the other hand, emission from the quantum state
$|c(t) \rangle$ has the extra contribution in the phase $\tilde{\Phi} (t) $ from
the internal ion energy $\propto \epsilon_{eg}$ which competes with
the orbital contribution giving rise to a kind of
non-linear resonance.

It is convenient for a comparison to introduce new angle variables
tangential to the circulating ion:
\begin{eqnarray}
&&
(\theta' , \psi' ) = (\theta + u, \psi)
\,,
\end{eqnarray}
such that $\theta' =0$ corresponds to the tangential direction
at different time $t = u\rho/\beta$.
The phase integral multiplied by the photon momentum phase space
is then
\begin{eqnarray}
&&
d\Omega \int_0^u dy \Phi(y; \theta', \psi) =
\int_0^u dy d\Omega' \Phi(y; \theta', \psi)
\,.
\end{eqnarray}
Keeping  in mind a photon extraction scheme into the outside of the ring,
one may take small angular regions $\propto \Delta \theta$ near the tangential direction of
$\theta' = 0$,
\begin{eqnarray}
&&
d\Omega' = \Delta \theta d\psi \cos \psi
\,, \hspace{0.5cm}
d\Omega \int_0^u dy \Phi(y; \theta' = 0, \psi)
\sim \Delta \theta d\psi \cos \psi
\int_0^u dy \Phi(y;  0, \psi)
\,.
\end{eqnarray}

\begin{figure*}[htbp]
 \begin{center}
 \epsfxsize=0.6\textwidth
 \centerline{\epsfbox{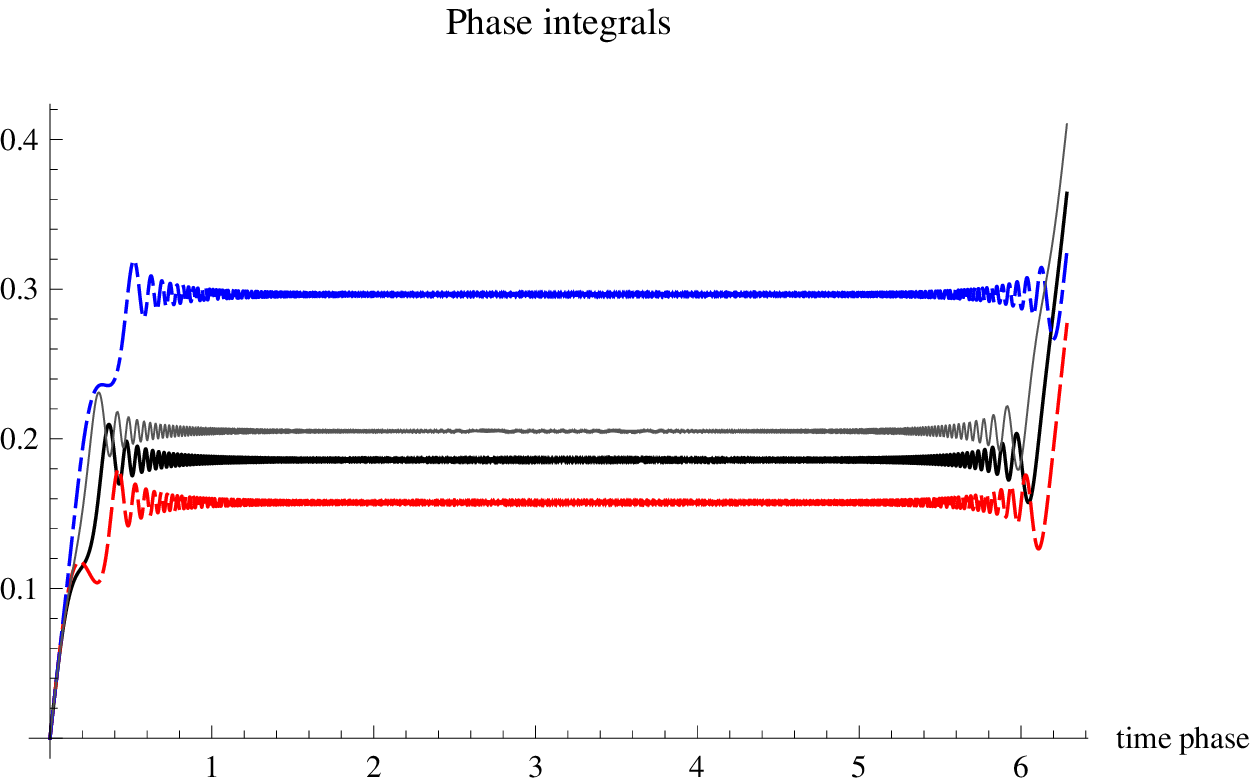}} \hspace*{\fill}
   \caption{Non-forward phase integral
$\int_0^u dy \cos \Phi(y; \theta') $ for a few values of
$\theta'$:
0.05 in solid black, 0.1 in dashed red, 0.2 in dash-dotted blue,
and 0 in dotted black.
Two period ion circulation of $u= 0\sim 4\pi$ were taken for the end point of time integral
without de-coherence.
Assumed parameters are $b=10^3, a= 1.01 b$
}
   \label {non-forward rate integral}
 \end{center} 
\end{figure*}

Written in terms of the new angular variable $\theta'$,
we compare two terms in eq.(\ref {phase 1}) , ($u = \beta t/\rho$)
\begin{eqnarray}
&&
\sin (\theta +u ) - \sin \theta = \sin \theta' + \sin (u - \theta')
= 
(1- \cos u) \sin \theta' + \sin u \cos\theta'
\,.
\label {tangential angles}
\end{eqnarray} 
Near the tangential direction of $\theta' = 0$, 
the first term in the last equality of eq.(\ref{tangential angles}) 
is small, both because of $\theta' \sim 0$ and a small phase $u$ region
contributing to the large rate.
In order to verify this assertion, we 
numerically simulated the phase integral keeping fixed the tangential
angle $\theta'$ at finite, non-vanishing values in eq.(\ref{tangential angles}).
The result is illustrated in Fig(\ref{non-forward rate integral}).
Simulations suggest that the phase integrals
for tangential angular regions of small, but finite
$|\theta'| < 0.1$ (solid and dotted black curves in 
Fig(\ref{non-forward rate integral})\,) agree well in $u < O(0.1)$.
Moreover, the agreement in the time phase region 
of nearly all $2\pi$ range except at points close to $2\pi$
is  good for smaller angle regions of $< O(0.05)$.
This phase region includes the most important
initial phases (regions up  around $\pi/4$),
smoothly matching to the stable plateau of large
phase integral.
It would be interesting to understand more deeply these behaviors
from the point of Floquet system described in Appendix B

In most of following discussions we shall suppress the angular $\theta'$ dependence
by fixing it at zero.
Thus, we shall use
\begin{eqnarray}
&&
\Phi(u; a, b) = b u - a \sin u
\,, \hspace{0.5cm}
a = \rho \omega \cos \psi
\,, \hspace{0.5cm}
b = \frac{\rho}{\beta} (\omega - \frac{\epsilon_{eg}}{\gamma}) 
\,,
\\ &&
\int_{0}^t dt'  \cos \tilde{\Phi} (t') \sim \frac{\rho}{\beta}
\int_0^{\beta t/\rho} d u \Phi(u; a, b)
\,.
\end{eqnarray}

Taking summation over photon polarization (helicity) gives
the differential spectrum of the form,
\begin{eqnarray}
&&
\frac{d^2\Gamma}{d\omega d\Omega} = \frac{2}{3(2\pi)^2}N d_{eg}^2
\epsilon_{eg}^2 \omega \gamma
\frac{\rho}{\beta} \int_0^{\beta t/\rho} du \cos \Phi(u)
\,, \hspace{0.5cm}
N = |\rho_{eg}(t)|^2 \frac{\rho I}{Q}  
\,,
\end{eqnarray}
with $|\rho_{eg}(t)| = \sin\theta_c \cos \theta_c$.
$N$ is the number of coherent ions available for
photon emission in the beam.

The radius $\rho$ of circular ion motion is of macroscopic length, and
one may take the infinite radius limit in the sense $\rho \epsilon_{eg} \gg 1$.
The largest contribution to the rates in the large radius
limit arises from 
the large region of $a, b$ in the phase function $\Phi(u)$.
Numerical result  for the  ion level spacing of 5 eV gives
\begin{eqnarray}
&&
\rho \epsilon_{eg} \sim 2.5 \times 10^{10} \frac{\rho}{ 1 {\rm km}}
\frac{\epsilon_{eg}}{5 {\rm eV}}
\,.
\end{eqnarray}

We consider detection of emitted photons after irradiating lasers from
counter-propagating directions.
The counter-propagating direction is chosen both to create a high coherence
and to utilize  laser frequencies boosted by the factor $2\gamma$
for the best match to deeper level spacing of ions.
Detectors are usually placed in a transport system
tangential to the ion circular motion.
The transport system for the photon extraction has a finite angular coverage $\Delta \theta$
in the ion beam plane,
and one should consider emitted photons from circulating ions 
in a limited time interval $\Delta t$, which is related to
the detected angular aperture $\Delta \theta$ by $\Delta t = \rho r\Delta \theta /v$,
$r$ being of order the ratio of the distance to
the detector to the circular radius to $R/\rho$.
The question now arises on where the extraction system
is to be placed in relation to the laser irradiation point.

It is shown in Appendix B that the Bessel function  is relevant to
the emission rate at one-period revolution $u = 2\pi$ of ion motion
after laser irradiation.
It  is however necessary to
calculate emission rates at any phase angle $u$ prior to one-period of circulation.
Consider for this purpose the phase integral $K(u)$,
\begin{eqnarray}
&&
K(u) = 
\int_0^{u} du' \cos (bu' - a \sin u')
\,.
\label {phase integral}
\end{eqnarray}
We are unaware of any simple analytic function that
gives this integral accurately.
Extensive numerical studies of this function for large $a, b$'s  
have been done accordingly.
A typical result is shown in Fig(\ref {approximate integral})
along with a truncated approximation to the third order in $u$
of the phase function.
There is a wide region of the phase $u$ giving a nearly flat plateau of integral
value at the half of the full integral, which is  $\pi J_b(a)$ (the Bessel function of
order $b$ and argument $a$).
For the parameter range $a \gg a - b >0$,
$\pi J_b(a) \sim \sqrt{\pi} (a^2 - b^2)^{-1/4}$.
See Appendix on some more details.
Stable photon emission rate is expected in this plateau region,
which is excellent for the purpose of extracting a large flux of photons as a  beam.
The initial behavior of emission rate at $u \ll 1$
shall be discussed in Section 4.

The differential spectrum rate in the stable extraction region is then given,
using a dimensionless energy $x=\omega/\epsilon_{eg}$, by
\begin{eqnarray}
&&
\frac{d^2 \Gamma}{ dx d\Omega} \sim
\frac{ A_{eg}}{2 \sqrt{\pi}} N 
\sqrt{\frac{\rho \epsilon_{eg} }{\beta} } \gamma 
 x  (\beta^2 x^2 \cos^2\psi \cos^2 \theta
- (x - \sqrt{1-\beta^2})^2)^{-1/4}
\,,
\label {differential spectrum}
\\ &&
x_- \leq x \leq x_+
\,, \hspace{0.5cm}
x_{\pm} = \sqrt{\frac{1 \pm \beta}{1 \mp \beta}}
\,,
\end{eqnarray}
with  the Einstein coefficient (decay rate) given by $A_{eg} = d_{eg}^2 \epsilon_{eg}^3/3\pi$.
This is our fundamental formula giving the angular distribution and the energy spectrum
of photon emission.
The overall rate factor is numerically given by
\begin{eqnarray}
&&
\frac{1 }{2 \sqrt{\pi}} A_{eg}N \sqrt{\rho \epsilon_{eg} } \sim
2.82 \times 10^{15} {\rm Hz} \frac{A_{eg} }{1 {\rm kHz}} \sqrt{\frac{\rho \epsilon_{eg} }{10^{10} }}
\frac{N}{10^8}
\,. 
\end{eqnarray}

\vspace{0.5cm}
\section
{\bf Energy spectrum at the forward
direction and  angular distribution}

In this section basic quantities relevant to detection of
extracted photons are calculated in the stable plateau region
of phases $< O(\pi/4)$.
Outside this region the approximation that neglects
the term $1- \cos u$ in eq.(\ref{tangential angles}) is questionable.
We first show the photon energy spectrum at the forward
direction.
For this purpose we integrate over the fundamental formula,
eq.(\ref{differential spectrum}), over a small solid angle area $\pi \Delta^2$
at the forward direction $\psi = \theta = 0$.
This can be done using
\begin{eqnarray}
&&
\int_{\sqrt{\psi^2 + \theta^2} \leq \Delta} d\psi d\theta \left(
A^2 - B^2 (\psi^2 + \theta^2)
\right)^{-1/4} = \frac{4\pi}{3} \frac{1}{B^2}
\left( A^{3/2} - (A^2 - B^2 \Delta^2)^{3/4}
\right)
\,.
\end{eqnarray}
Taking $A^2, B^2$ relevant  to our problem and
expanding in terms of the small angle factor $\Delta^2$,
one finds that 
\begin{eqnarray}
&&
\int_{\sqrt{\psi^2 + \theta^2} \leq \Delta} d\psi d\theta 
(\beta^2 x^2 \cos^2\psi \cos^2 \theta
- (x - \sqrt{1-\beta^2})^2)^{-1/4}
\nonumber
\\ &&
\sim
\pi \Delta^2 \gamma^{1/2}
\left( (x - x_-) (x_+ -x )
\right)^{-1/4}
\,.
\end{eqnarray}
This formula is valid in a limited  region of $\Delta^2$,
which gives  a $\Delta$ dependent range  of allowed photon energies;
\begin{eqnarray}
&&
\frac{1}{ 1+ \beta^2 \gamma^2 \Delta^2}
(\gamma - \sqrt{\gamma^2 ( 1- \beta^2\Delta^2)-1 })
\leq x \leq
\frac{1}{ 1+ \beta^2 \gamma^2 \Delta^2}
(\gamma + \sqrt{\gamma^2 ( 1- \beta^2\Delta^2) -1 })
\,.
\end{eqnarray}
The forward energy spectrum rate is given by
either of the following two forms,
\begin{eqnarray}
&&
\left(\frac{d\Gamma}{dx} \right)_0 = N \pi \Delta^2
\frac{A_{eg}}{2\sqrt{\pi}} \sqrt{\rho \epsilon_{eg}}
\frac{\gamma^{3/2}}{\sqrt{\beta}}
x \left( (x - x_-) (x_+ -x )
\right)^{-1/4}
\,,
\label {forward energy spectrum}
\\ &&
\left(\frac{d\Gamma}{dy} \right)_0 \rightarrow N \pi \Delta^2
\frac{A_{eg}}{\sqrt{2\pi}} \sqrt{\rho \epsilon_{eg}}
\gamma^{2} y^{3/4} (1-y)^{-1/4}
\,, \hspace{0.5cm} 
y = \frac{ \omega} {\omega_{{\rm max}}} \sim 
\frac{ \omega} {2\gamma \epsilon_{eg}} 
\,, \hspace{0.5cm}
{\rm as \;} \gamma \rightarrow \infty
\,.
\end{eqnarray}

\begin{figure*}[htbp]
 \begin{center}
 \epsfxsize=0.6\textwidth
 \centerline{\epsfbox{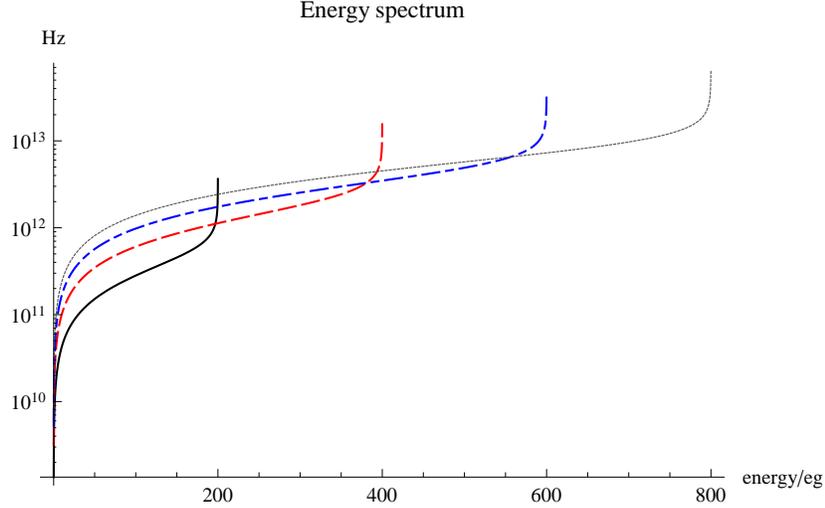}} \hspace*{\fill}
   \caption{
Photon energy spectrum per unit solid angle area (eq.(\ref {forward energy spectrum}) divided by $\pi \Delta^2$)
at the forward direction
for a few choices of the boost factor:
$\gamma =$ 100 in solid black, 200 in dashed red,  300 in dash-dotted blue,
and 400 in dotted black.
The angular resolution of $\Delta = 0.01/\gamma$ was taken here.
Other assumed parameters are $A_{eg} = $1kHz, $\rho \epsilon_{eg} = 10^{10}, N = 10^8$
and rates scale as $\propto A_{eg} \sqrt{ \rho \epsilon_{eg}}N$.
}
\label {gamma spectrum 2}
 \end{center} 
\end{figure*}

We expect under normal experimental circumstances that $\Delta \leq O(1/\gamma)$.
In Fig(\ref{gamma spectrum 2}) we illustrate 
the forward energy spectrum per unit solid angle area $\pi \Delta^2 = 1$.
The Jacobian peak at the highest energy is
clearly visible for smaller values of the angular resolution
$\Delta$.
The peak degrades when the angular coverage $\Delta$ becomes larger,
for instance at $\Delta \geq O(0.1)/\gamma$.
The Jacobian peak suggests a high degree of
a correlation between the photon energy and
its emission angle.

\begin{figure}[htbp]
\begin{center}
\begin{minipage}{7.5cm}
 \epsfxsize=1\textwidth
 \centerline{\epsfbox{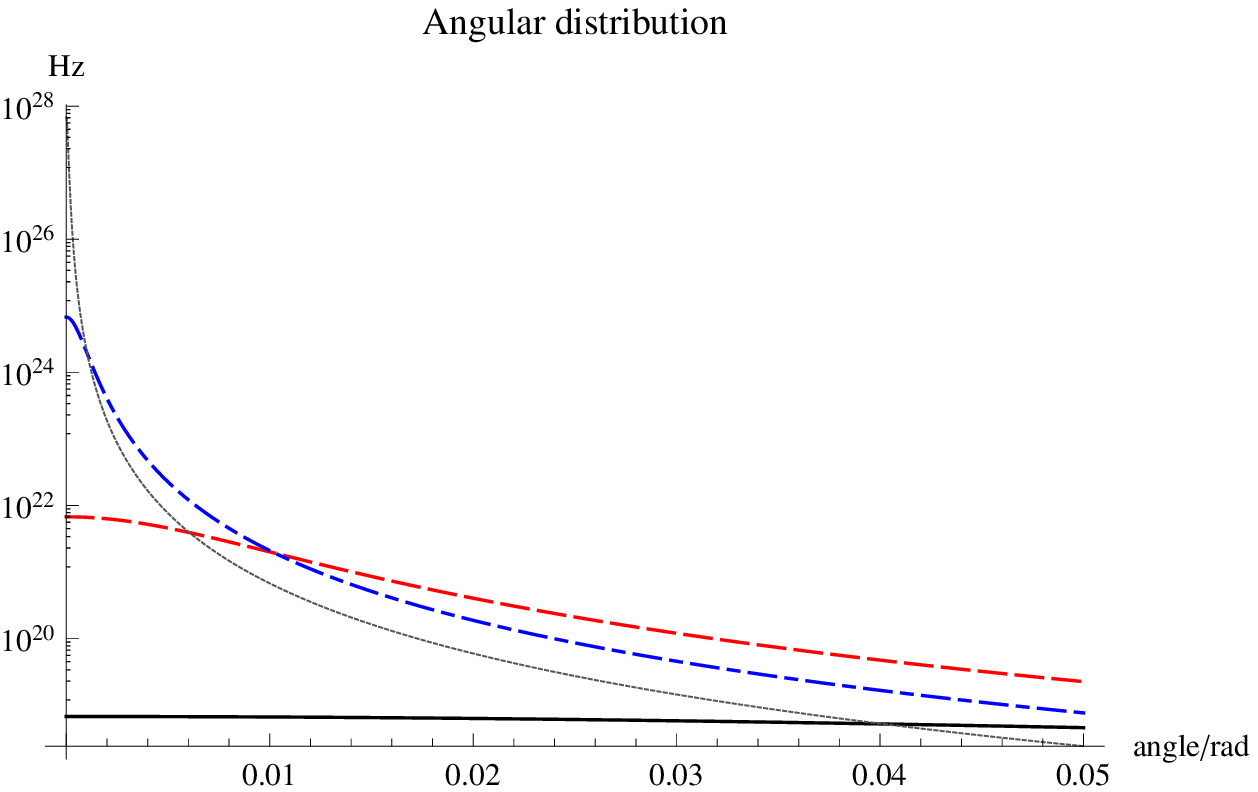}} \hspace*{\fill}
\end{minipage}
\begin{minipage}{0.5cm}
$\;$
\end{minipage}
\begin{minipage}{7.5cm}
   \epsfxsize=1\textwidth
\centerline{\epsfbox{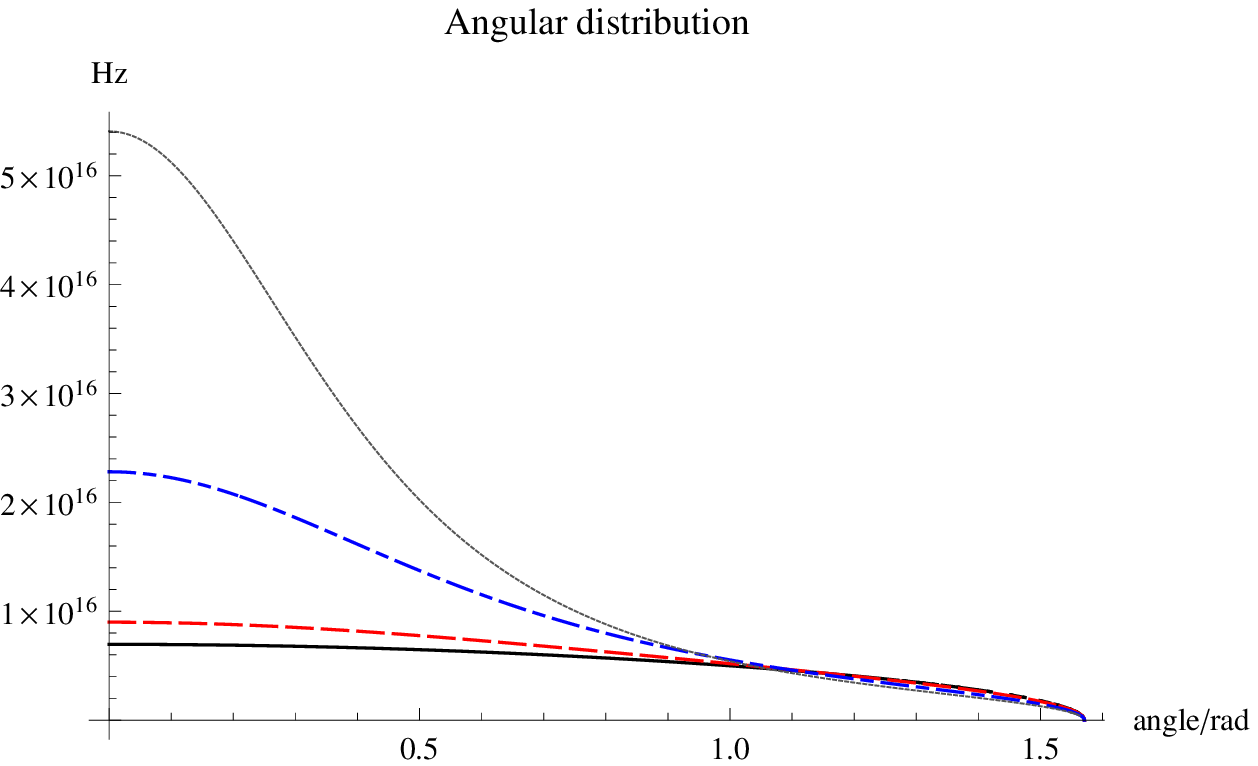}} \hspace*{\fill}
\end{minipage}
\caption{Angular distribution
either at $\psi=0$ or $\theta=0$  in the single photon emission  for a few choices of
the boost factor.
In the left panel
$\gamma =$ 10 in solid black, 100 in dashed red,  1000 in dash-dotted blue,
and $10^4$ in dotted black.
In the right panel
$\gamma =$ 1.01 in solid black, 1.1 in dashed red,  1.5 in dash-dotted blue,
and 2 in dotted black.
Other assumed parameters are $A_{eg} = $1kHz, $\rho \epsilon_{eg} = 10^{10}, N = 10^8$
and rates scale as $\propto A_{eg} \sqrt{ \rho \epsilon_{eg}}N$.
}
\label{1g angular dist}
\end{center}
\end{figure}

We next calculate the angular distribution after the photon energy integration.
The formula of angular integration gives
\begin{eqnarray}
&&
\frac{d\Gamma}{d\Omega} = \frac{1}{2\sqrt{\pi}} N A_{eg} \sqrt{\rho \epsilon_{eg}}
\frac{\gamma}{\sqrt{\beta}} 
\int_{X_-}^{X_+} dx \frac{x}{(\beta^2 x^2 \cos^2 \psi \cos^2 \theta - (x - \sqrt{1-\beta^2})^2)^{1/4} }
\\ &&
= c_0 N  A_{eg} \sqrt{\rho \epsilon_{eg}} \gamma^{-1/2} 
\frac{ \sqrt{\cos \psi \cos \theta } }{(1- \beta^2 \cos^2 \psi \cos^2 \theta)^{7/4} }
\rightarrow c_0  NA_{eg} \sqrt{\rho \epsilon_{eg}} \gamma^{3} 
\frac{1} {(1- \gamma^2 (\psi^2 + \theta^2)\,)^{7/4} }
\,,
\\ &&
X_{\pm} = \frac{1}{\gamma (1 \mp \beta \cos \psi \cos \theta)}
\,, \hspace{0.5cm}
c_0 = 2 \frac{ \Gamma(\frac{3}{4})} {  \Gamma(\frac{1}{4})} \sim 0.676
\,.
\end{eqnarray}
This angular distribution is illustrated in Fig(\ref{1g angular dist}).
The forward peaking as the boost factor increases is clearly
observed already at intermediate $\gamma$ values.

Finally, we calculate the total photon emission rate
by integrating both energy and angle variables.
The result is given by 
\begin{eqnarray}
&&
\Gamma \sim 2.8  N A_{eg} \sqrt{\rho \epsilon_{eg}}
\gamma
\,,
\label {total rate 1}
\end{eqnarray}
which holds in both large $\gamma$ and small $\beta$ limits.

We observe that compared to the spontaneous emission rate $A_{eg} N$,
the quantum ion beam gives rise to a rate enhanced by a factor $\propto \sqrt{\rho\epsilon_{eg}} $,
except the factor $\propto \gamma$.
Since $\rho$ is of a macroscopic size and is much larger than
the microscopic scale $1/\epsilon_{eg}$, this may become gigantic.

A non-trivial aspect of $\gamma-$dependence should be noted for
the collimated photon emission.
For a large $\gamma$ the angular coverage is very much limited to an
angular area of order $1/\gamma^2$ and the total rate is diminished by
this coverage factor.
This means that in a unit solid angle area in the forward narrow cone
the differential angular rate is effectively of the order $\gamma^2$ larger than
the total rate.

Calculations so far presented are exact except
the use of approximate form of large order Bessel function.
An alternative method of total rate calculation is to treat
the angular variables $\theta, \psi$ symmetrically, and to
approximate the product function
\begin{eqnarray}
&&
\cos^2 \psi \cos^2 \theta \sim 1 - (\psi^2 + \theta^2)
\,,
\end{eqnarray}
which is valid at large $\gamma$'s showing the highly collimated angular distribution.
This method can readily be extended to
the case of multiple particle emission such as the neutrino pair emission.
To make the behavior of the rates in the high energy limit ($\gamma \rightarrow \infty$) more transparent,
it is  convenient to use the energy rescaled by
the maximum  energy $x_+ \epsilon_{eg}$ and to introduce the variable $y$ defined below.
Resulting rates are as follows:
\begin{eqnarray}
&&
\frac{d\Gamma'}{dy} = c_3 A_{eg} N \sqrt{\rho \epsilon_{eg}} G(y)
\,, \hspace{0.5cm}
c_3 = \frac{4 \sqrt{2\pi}}{3} \sim 3.34
\,,\hspace{0.5cm}
 y = \frac{ x}{x_+}
\,, \hspace{0.5cm}
y_-\leq y \leq 1
\,, \hspace{0.5cm}
y_-=  \frac{1-\beta}{1+\beta}
\,, 
\\ &&
G(y) = \frac{ \left( (y - y_- )( 1- y)\right)^{3/4}}{ y} 
\,.
\end{eqnarray}
A more useful approximation to treat the total rate
at any velocity is given by
\begin{eqnarray}
&&
\Gamma' = c_4 A_{eg} N \sqrt{\rho \epsilon_{eg}} \gamma
\,,  \hspace{0.5cm}
c_4 = \frac{4 \sqrt{2\pi}}{3} B(\frac{3}{4}, \frac{7}{4}) \sim 2.83
\,.
\label {total rate approximate}
\end{eqnarray}
The agreement of the parameter dependence and an excellent closeness of
constants in eq..(\ref{total rate 1}) and eq.(\ref{total rate approximate})
gives a confidence in the approximation here.

\vspace{0.5cm} 
\section
{\bf Initial phase of photon emission and  question of de-coherence}

The photon emission discussed so far is based on
a high coherence of quantum ion beam, but
it may decay by emitting  photons continuously.
We shall first discuss the coherence decay law in general,
and then address the important question of how
the coherence is maintained at initial phases.

Using the total emission rate for a single ion $\Gamma_s(t)$,
one derives the decay law from
\begin{eqnarray}
&&
\frac{d\rho_{eg}}{dt} = - \frac{\Gamma_s}{2} \rho_{eg}
\,.
\end{eqnarray}
It is solved as
\begin{eqnarray}
&&
\rho_{eg}(t)  = \rho_{eg}(t_0) \exp[-\int_{t_0}^t \frac{\Gamma_s(t')}{2}dt']
\,.
\end{eqnarray}
Note that for time dependent $\Gamma_s(t)$, the decay is not of the usual exponential form.
This coherence decay is accompanied by the population decay
described by
\begin{eqnarray}
&&
\frac{d\rho_{ee}}{dt} = -\Gamma_s \rho_{ee}  
\,.
\end{eqnarray}

\begin{figure*}[htbp]
 \begin{center}
 \epsfxsize=0.6\textwidth
 \centerline{\epsfbox{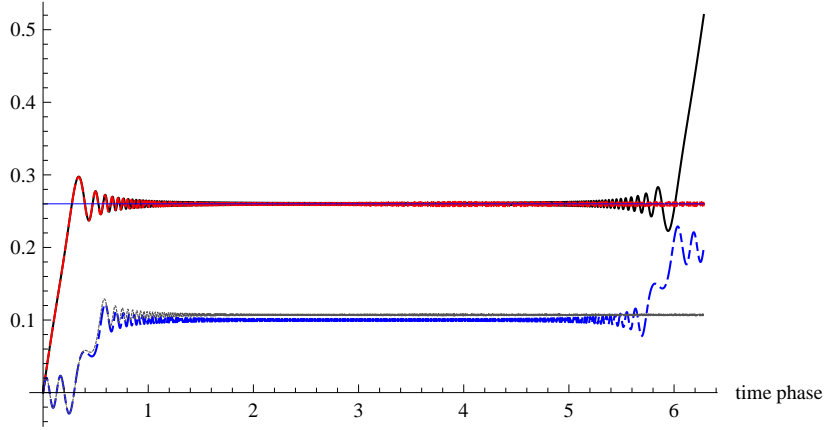}} \hspace*{\fill}
   \caption{
Approximate phase integral by truncation to the third order
compared with exact results.
exact case of $(b,a) = 500(1,1.01)$ in solid black,  its approximate
case in dashed red, and exact case of $(b,a) = 500(1,1.1)$ in dash-dotted blue,
and its approximate case in dotted black.
The straight line shows the $J_b(a)$ value for  $(b,a) = 500(1,1.01)$.
}
   \label {approximate integral}
 \end{center} 
\end{figure*}

One may use a truncated time expansion in
the initial and intermediate phase region for
estimate of the  crucial phase integral.
We thus expand the phase function $\Phi(u)$ in terms of the circulating phase 
variable $u$ to its third order:
\begin{eqnarray}
&&
\Phi(u) \sim -(a-b) u + \frac{a}{6} u^3
\,.
\end{eqnarray}
This approximation is compared with the exact phase integral
in Fig(\ref {approximate integral}).
It is thus clear that the third order approximation
is excellent except in a small region near the returning phase point of
$u= 2\pi$.

For further analytic calculations it
is important to locate stationary points of the phase integral.
The stationary phase point given by
the vanishing derivative $\Phi'(u) = 0$
exists at
\begin{eqnarray}
&&
u = u_{\pm} = \pm \sqrt{\frac{2(a-b) } {a }} \,,
\hspace{0.3cm} {\rm for \;} \frac{a-b } {a } \geq 0
\,.
\end{eqnarray}
The Gaussian phase approximation to this function
and the resulting phase integral  gives
\begin{eqnarray}
&&
\Phi(u) \sim
- \frac{2\sqrt{2}} {3} (a-b) \sqrt{ \frac{ a-b}{ a}}
+ \sqrt{\frac{a(a-b) }{ 2}} (u - \sqrt{ \frac{2( a-b)}{ a}})^2
\,,
\\ &&
\int_0^u dy \Phi(y) \sim
\left( \frac{2}{ a(a-b)}\right)^{1/4}
\Re \left( e^{-iX} \int_0^Y dz \exp[i (z - W)^2]
\right) 
\,,
\\ &&
X = \frac{ 2 \sqrt{2}} {3} (a -b) \sqrt{\frac{a-b }{a }}
\,, \hspace{0.5cm}
Y = u \left( \frac{ a(a-b)}{2}
\right)^{1/4}
\,,\hspace{0.5cm}
W = b \left( \frac {2} { a(a-b)}
\right)^{1/4}
\,.
\end{eqnarray}
In the limit of $Y \gg W$,
one may replace $Y \rightarrow \infty\,, W \rightarrow 0$
to derive a Fresnel type of integral in an infinite range,
\begin{eqnarray}
&&
\int_0^{\infty} dy \Phi(y) \sim \left( \frac {2} { a(a-b)}
\right)^{1/4}
\frac{\sqrt{\pi}}{2} \cos (X - \frac{\pi}{4})
\,.
\end{eqnarray}

Inserting relevant quantities for $a,b$ gives
\begin{eqnarray}
&&
\frac{ d^2 \Gamma}{dx d\Omega} = \frac{1}{2^{9/2}\sqrt{\pi}} A_{eg} N \sqrt{\rho \epsilon_{eg}} 
\frac{ (1+\beta)^{1/4}}{ \beta^{3/4}} \gamma^{3/2} F_{1g}(x)
\,,
\\ &&
F_{1g}(x) = 
x \left( x (x_+ - x) - \frac{\beta ( 1+ \beta)}{2} \gamma^2 x^2 (\theta^2 + \phi^2)
 \right)^{-1/4}
\nonumber \\ && \times
\cos
\left( \frac{\rho \epsilon_{eg}}{3} (\sqrt{\beta(1+\beta)}\gamma)^{-3} 
x \left(
\frac{ x_+ -x}{x} -  \frac{\beta ( 1+ \beta)}{2} \gamma^2  (\theta^2 + \phi^2)
\right)^{3/2}
- \frac{\pi}{4}
\right)
\,.
\label {oscillatory spectrum}
\end{eqnarray}
The major difference from the previous formula (\ref {differential spectrum}),
in particular at large $\gamma$'s,
is the presence of the oscillating function having an interesting combination of large
factors, $\rho \epsilon_{eg}/\gamma^2$ with $x \propto \gamma$;
all other factors are in reasonable agreement with the previous result.

Another difference is the low $\beta$ behavior,
arising from $\propto \left( a(a-b) \right)^{-1/4}$,
which should be compared with the previous
approximation from the Bessel function giving
$\propto \left( a^2-b^2 \right)^{-1/4}$.
Although the difference at large $\gamma$'s is
minor, the low  $\beta$ behavior is quite different.
Indeed, the stationary phase approximation here
cannot reproduce the correct behavior
$\propto \sqrt{\beta}$ in the total rate.

This formula is valid for the time or its related phase domain,
\begin{eqnarray}
&&
t > t_*
\,, \hspace{0.5cm}
 t_* = \rho \sqrt{\frac{ 2}{ \beta ( 1+ \beta)}} \frac{1}{\beta \gamma} \left(
\frac{ x_+ -x}{x} -  \frac{\beta ( 1+ \beta)}{2} \gamma^2  (\theta^2 + \psi^2)
\right)^{1/2}
\end{eqnarray}
Due to the oscillating factor the major contribution arises from
the photon phase space region of
\begin{eqnarray}
&&
\frac{ x}{x_+} \left(
\frac{ x_+ -x}{x} -  \frac{\beta ( 1+ \beta)}{2} \gamma^2  (\theta^2 + \psi^2)
\right)^{3/2} < \frac{ 3} {\rho \epsilon_{eg} }\beta^{3/2} (1+\beta)^{1/2} \gamma^2
\,.
\end{eqnarray} 
The combination of parameters that appear here is of order,
\begin{eqnarray}
&&
\frac{ \gamma^2} {\rho \epsilon_{eg} } = 10^{-4} (\frac{ \gamma}{10^3 })^2 \frac{ 10^{10}}{\rho \epsilon_{eg} }
\,.
\end{eqnarray}

The physical implication of this result is that
in earlier phases of circulations ions tend to emit photons
into more restricted region of their phase space than
the stable plateau region.
Clarification of 
how this transient region is smoothly connected to
the stable phase region of the plateau needs more
refined treatment.

\begin{figure*}[htbp]
 \begin{center}
\epsfxsize=0.6\textwidth
 \centerline{\epsfbox{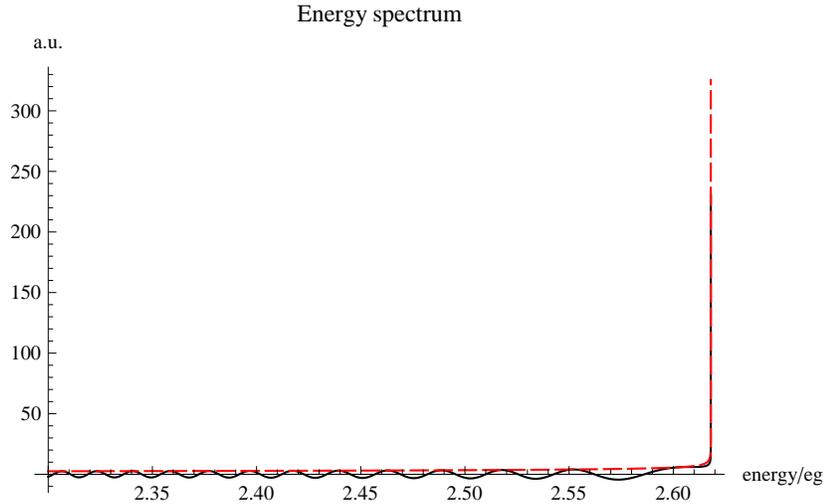}} \hspace*{\fill}
   \caption{
Effect of oscillatory behavior in the forward energy spectrum.
Formula given by eq.(\ref{oscillatory spectrum}) with (in solid black)
and without (in dashed red) the sinusoidal function.
Assumed parameters are $\gamma = 1.5, \rho \epsilon_{eg} = 10^{14}$
case in dotted black.
}
   \label {spectrum oscillation}
 \end{center} 
\end{figure*}

At the quantitative level
the effect of oscillatory term is important only
at low $\gamma$'s.
We illustrate an example in Fig(\ref{spectrum oscillation}),
which shows that the oscillation effect is not significant
at energies contributing to large rates.

From the intermediate time behavior of emission rate
just given, one can estimate the e-folding factor in
the coherence decay law,
\begin{eqnarray}
&&
\Gamma_s =2.8 A_{eg} \sqrt{\rho \epsilon_{eg}} \gamma
\,,
\\ &&
\frac{\Gamma_s L }{2} \sim 5.6 \times 10^{-2} \frac{ A_{eg}}{ {\rm Hz}} \frac{L }{10 {\rm m} } 
\frac{ \gamma}{ 10^3} \sqrt{  \frac{ \rho \epsilon_{eg}}{ 10^{10}}}
\,.
\end{eqnarray}
This formula is valid only for $t> \langle t_* \rangle$ which is an averaged limit time
estimated as an average over energies at the forward direction;
\begin{eqnarray}
&&
\langle t_* \rangle \sim \sqrt{2}\pi \frac{\rho}{\beta^{3/2} \sqrt{1+\beta} \gamma}
\,.
\end{eqnarray}

At earlier times of $t <  \langle t_* \rangle$ one may attempt to use the linearized formula
of emission rate,
\begin{eqnarray}
&&
\Gamma_s(t) = 2.8 A_{eg} \sqrt{\rho \epsilon_{eg}} \gamma
\frac{t}{\langle t_* \rangle} 
\,.
\end{eqnarray}
This consideration leads to the initial decay law of the form,
\begin{eqnarray}
&&
\rho_{eg}(t) = \rho_{eg}(0)
\exp[- (\frac{t}{t_0})^2]
\,, \hspace{0.5cm}
t_0 \sim 1.2 {\rm m} \frac{ 10^3}{\gamma }\sqrt{  \frac{ {\rm Hz}}{A_{eg} }}\sqrt{  \frac{\rho }{{\rm km} }}
(\frac{ \rho \epsilon_{eg}} {10^{10}})^{-1/4}
\,.
\end{eqnarray}

It would be nice if one can make more refined analysis
in the initial phases of time evolution,
because the approximation $\propto t$ for the initial time behavior
is too crude..

\vspace{0.5cm} 
\section
{\bf  Neutrino pair  emission rates}

The cases of multiple particle emission, 
in particular a neutrino pair,
are of great interest.
Our formalism may be directly extended to these cases,
which we shall turn to.

For simplicity we shall take massless neutrinos of three flavors,
which is an excellent approximation for neutrino energies $E_i$
much larger than their masses.
Quantities that appear in the phase integral are
changed to
\begin{eqnarray}
&&
\Phi(u) = bu - a\sin u
\,, \hspace{0.5cm}
b = 
\frac{\rho}{\beta} (E_1 + E_2 - \frac{\epsilon_{eg}}{\gamma})
\,, \hspace{0.5cm}
a = 
\rho \left(E_1 \cos\psi_1 \cos \theta_1
+ E_2 \cos\psi_2 \cos \theta_2 \right)
\,.
\end{eqnarray}
Matrix element factors are readily worked out
by squaring the electron spin transition moment $\vec{S}_e = \langle g|\vec{\sigma}|e \rangle/2$ arising
from the axial vector part of four Fermi interaction as in \cite{pair beam}.
The calculated differential rate for three neutrino flavor
pairs is
\begin{eqnarray}
&&
\frac{ d^4 \Gamma_{2\nu}}{dy_1 dy_2 d\Omega_1 d\Omega_2} =
\frac{ \sqrt{\pi}}{16 (2\pi)^6} G_F^2 \epsilon_{eg}^5 N \sqrt{\rho \epsilon_{eg}} 
\vec{S}_e^2 ( 1+ \frac{2}{3} \beta^2 \gamma^2)
\frac{1}{\sqrt{\beta}\gamma}x_+^6
\nonumber 
\\ &&
\times
y_1^2 y_2^2 \left( 1 + \frac{1}{3}\cos \psi_1 \cos\psi_2 \cos(\theta_1- \theta_2) + \frac{1}{3}\sin \psi_1 \sin \psi_2
\right)
\nonumber \\ &&
\times
\left( \beta^2 (y_1 \cos\psi_1\cos\theta_1 + y_2 \cos\psi_2\cos\theta_2)^2 - (y_1 + y_2 - \frac{1}{\gamma x_+})^2
\right)^{-1/4}
\,,
\label {differential 2n rates}
\end{eqnarray}
with $x_+ = \sqrt{(1+\beta)/(1-\beta)} \sim 2 \gamma$
and $y_i = E_i/(x_+ \epsilon_{eg})$.
After angular integrations that give $O(\gamma^{-4})$ factor,
one may derive a formula of the total pair emission rate,
\begin{eqnarray}
&&
\Gamma_{2\nu} \sim 
c_4  G_F^2 \epsilon_{eg}^5 N \sqrt{\rho \epsilon_{eg}} 
\vec{S}_e^2 ( 1+ \frac{2}{3} \beta^2 \gamma^2)
\beta^{-9/2} (1+\beta)^{11/2} \gamma I
\,,
\label {total 2n rate}
\\ &&
I =
\int dy_1 dy_2 \frac{y_1 y_2}{(y_1 + y_2)^2} ((y_1 + y_2 - y_-) (1 - y_1- y_2)\,)^{7/4}
\,,
\\ &&
c_4 = \frac{\sqrt{\pi}}{12 (2\pi)^6} \int_{| \vec{x} | \leq 1} dv_4 (1 - \vec{x}^2 )^{-1/4}
 = \frac{1}{128 (2\pi)^3} \frac{\Gamma(3/4)}{\Gamma(13/4)} 
\sim 1.21 \times 10^{-4}
\,, 
\\ &&
c_4  G_F^2 \epsilon_{eg}^5 N \sqrt{\rho \epsilon_{eg}} 
\sim 2.5  {\rm Hz} \sqrt{ \frac{\rho \epsilon_{eg}}{10^{14}}}\frac{N}{10^8}
(\frac{ \epsilon_{eg}}{10 {\rm keV}})^5
\,.
\label {overall 2n rate}
\end{eqnarray}
The integral related to the constant $c_4$ is over the 4d volume of radius unity.
For this estimate we took a large radius $\rho$ of ion circular motion and
the level spacing $\epsilon_{eg}$ appropriate for
high energy neutrino pair production.

Dependence of the total  rate $\propto \gamma^3$ 
of eq.(\ref{total 2n rate}) in the high energy limit is different 
from the $\propto \gamma^4$ in \cite{pair beam}.
The difference is traced to a different angular distribution
in  the stationary phase approximation.
In \cite{pair beam} the angular $\theta, \psi$ distribution
is asymmetric, and it has a wider range of allowed angles in $\theta_i$ variables:
only the relative opening angle $\theta_1 - \theta_2$ is limited
by $1/\gamma$, but their individuals $\theta_i$ are not limited
by this factor.
This does not give an extra $1/\gamma$ suppression
in the result of the total rate, which explains the
difference from the present work.
As to the total rate we believe that the present method
is closer to the correct, more precise result.
But it is important to note that
what is to be compared with actual observations is
the rate within a given aperture of angles, and
this should be calculated more precisely by
taking into account the geometry of detector system.
This way one may obtain an effective enhancement
of a $\gamma$ power.

\begin{figure*}[htbp]
 \begin{center}
 \epsfxsize=0.6\textwidth
 \centerline{\epsfbox{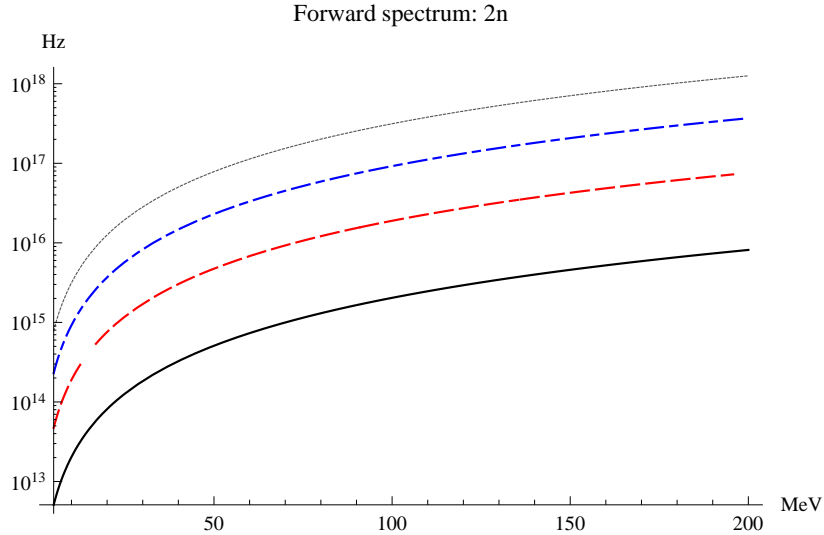}} \hspace*{\fill}
   \caption{Neutrino spectrum rate at the forward direction in  neutrino pair emission,
assuming the overall factor of 0.05 Hz in eq.(\ref{n-spectrum}).
Chosen $\gamma$ factors are 4000 in solid black, 6000 in dashed red,
8000 in dash-dotted blue, and $10^4$ in dotted black.
}
\label {n-spectrum rate}
 \end{center} 
\end{figure*}

In Fig(\ref {n-spectrum rate}) we illustrate 
the forward spectrum rate of a single neutrino in the
pair production process.
The basic formula
derived by taking $\psi_i = \theta_i = 0$ in eq.(\ref{differential 2n rates})
 is given in terms of $y= x \sqrt{(1+\beta)/(1-\beta) }$,
\begin{eqnarray}
&&
S(y) = 
0.05 {\rm Hz} \sqrt{\frac{\rho \epsilon_{eg} }{10^{14} }} \frac{N}{10^8}
\vec{S}_e^2 ( 1 + \frac{2}{3} \beta^2 \gamma^2) \frac{(1+\beta)^6}{\sqrt{\beta} }\gamma^{11/2}
\nonumber
\\ &&
\times
\int dy_2 y^2 y_2^2 \left( 
(y_1 + y_2 - y_-) (y_+ - y_1 -y_2)
\right)^{-1/4}
\,,
\label {n-spectrum}
\end{eqnarray}
which is valid for rates per unit solid angle area at the forward
direction of $\theta = \psi = 0$.
The spin factor $\vec{S}_e^2 =1$ and $N= 10^8, \rho \epsilon_{eg} = 10^{14}$ were taken for simplicity.

Now we would like to discuss 
a  possible obstacle against detection of
the neutrino pair emission.
Gigantic backgrounds of QED processes are not
really a problem,
because all emitted photons are beam-dumped by some
experimental facility before extracted neutrino pair beam
is used for experiments.
The important question is whether the absolute neutrino
pair emission is large enough for experiments away from
the ion ring.
Our calculations here show that
when the circulating ion and the accelerator parameter
are appropriately chosen, this is not a problem.

\vspace{0.5cm}
In summary,
we presented both the differential and the total rates
of produced single photon and neutrino-pair
emitted from quantum ions in circular motion.
Results were given for any velocity of circulating ion,
hence should be useful for experimental investigations
of these processes.
An important constraint on parameters of
ions and circulation ring was derived to
ensure the initial coherence necessary
for the process.
In a sequel to the present work \cite{asaka et al}
we shall discuss how neutrino
oscillation experiments can measure
important parameters of neutrino properties.

\vspace{0.5cm} 
\section
{\bf Appendix A: Rates of two-photon emission}

Here we calculate rates of two-photon emission
and discuss which of the single or the two-photon
emission is dominant for large boost values relevant
to the large rate region of the neutrino pair emission.

Differential spectrum of two-photon emission  is calculated as
\begin{eqnarray}
&&
\frac{d^4 \Gamma_{2\gamma}}{dy_1dy_2 d\Omega_1d\Omega_2}
= \frac{\sqrt{\pi}}{4(2\pi)^4}N
\frac{ A_{pe}A_{pg}\epsilon_{eg}}{\epsilon_{pe} \epsilon_{pg}}\epsilon_{eg}
\sqrt{\frac{ \rho \epsilon_{eg}}{\beta}}(1+\beta)^6
\gamma^{4} y_1 y_2 M (y_1, y_2)
\nonumber 
\\ &&
\times
 (\beta^2 \gamma^2 ( y_1 \cos \psi_1 \cos \theta_1 + y_2 \cos \psi_2 \cos \theta_2)^2
- (\, \gamma( y_1 + y_2) - \frac{1}{x_+})^2)^{-1/4}
\,,
\\ &&
\hspace*{-0.5cm}
M  (y_1, y_2) = \frac{1}{( y_1 + \epsilon_{pe}/(x_+ \epsilon_{eg})\,)^2} + \frac{1}{( y_2 +\epsilon_{pe}/(x_+ \epsilon_{eg})\,)^2} 
+ \frac{3}{4} \frac{1}{( y_1 + \epsilon_{pe}/(x_+ \epsilon_{eg})\,) ( y_2 + \epsilon_{pe}/(x_+ \epsilon_{eg})\,)}
\,,
\end{eqnarray}
with $y_i = \omega_i/(x_+\epsilon_{eg})$.
The function $M$ is the squared sum of energy denominator factors
in the second order of perturbation theory.
In this estimate of matrix element $M$ we
replaced a factor $(\vec{k}_1 \cdot \vec{k}_2/\omega_1 \omega_2)^2$ in the interference term
by its average $1/2$, which is not precise, but
it would serve for our crude estimate.

The angular integrations over those of two photons
may be explicitly done, using the
expansion like $\cos \theta_i \sim 1 - \theta_i^2/2$,
which should be valid for large boosts.
The result after angular integrations is
\begin{eqnarray}
&&
\int d\Omega_1 \int d\Omega_2 
\left(
(y_1+ y_2 - y_-)(1 - y_1 - y_2) -\beta^2 \gamma^2 (y_1+ y_2)
(y_1 ( \psi_1^2 + \theta_1^2) + y_2 ( \psi_2^2 + \theta_2^2) \,
\right)^{-1/4}
\nonumber
\\ &&
\sim \frac{128}{231} \pi^2
 \frac{\Gamma(3/4)}{\Gamma(13/4)} 
(\beta \gamma)^{-4}
\frac{ \left( (y_1+ y_2 - y_-)(1 - y_1 - y_2)\right)^{7/4}}{y_1 y_2 (y_1 + y_2)^2}
\,.
\end{eqnarray}
The single photon spectrum shape after the second photon energy is integrated out
is given by
\begin{eqnarray}
&&
\frac{d\Gamma_{2\gamma}}{dy} = 
3.6 \times 10^{9} {\rm Hz}
\frac{ {\rm keV} \epsilon_{eg}}{\epsilon_{pe} \epsilon_{pg}}
\frac{ A_{pe}A_{pg}} {({\rm 10 MHz})^2}
\sqrt{ \frac{\rho \epsilon_{eg}}{10^{10}}}\frac{N}{10^8}
\frac{(1+\beta)^2}{\beta^{9/2}}
F_{2\gamma}(y)
\,,
\\ &&
F_{2\gamma}(y) = \int dy_2 (\frac{1}{y+ y_2})^2 \left(
(y+y_2 - y_-)(1 - y- y_2)
\right)^{7/4} M(y,y_2)
\,, \hspace{0.5cm}
y_- = \frac{1- \beta}{1+\beta}
\,.
\end{eqnarray}
with $x_+ = \sqrt{(1+\beta)/(1-\beta)}$.

\begin{figure*}[htbp]
 \begin{center}
 \epsfxsize=0.6\textwidth
 \centerline{\epsfbox{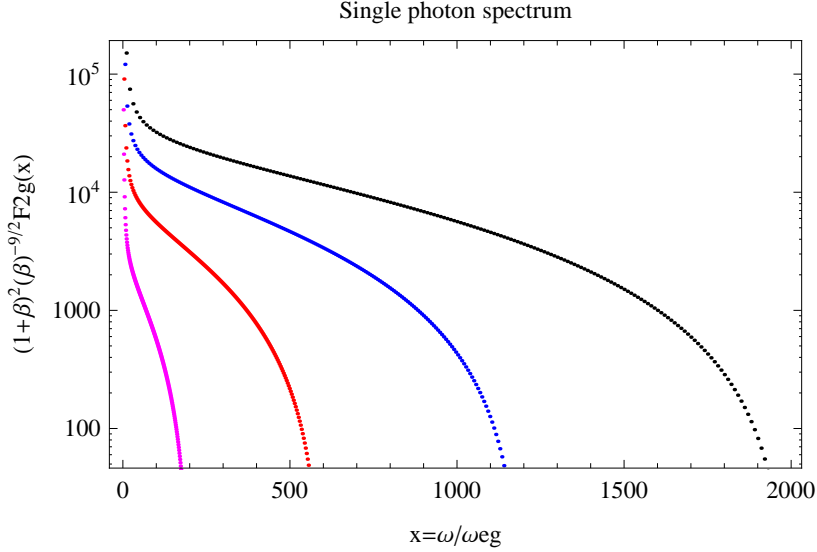}} \hspace*{\fill}
   \caption {
Energy spectrum shape of single photon
using $x= \omega/\epsilon_{eg}$
in two-photon emission for
$\gamma=$ 100 in magenda, 300 in red, 600 in blue,
and 1000 in black.
$\epsilon_{pe}/\epsilon_{eg} = 0.5$ is assumed.
}
\label {2g energy spectrum1}
 \end{center} 
\end{figure*}

Numerical results of the energy spectrum of a single photon in the two-photon
emission process are
illustrated in Fig(\ref{2g energy spectrum1})
for some values of the boost factor.
The end point of energy spectrum increases with $\gamma$
accompanied by rate increase.
In the high energy limit of $\gamma \gg 1$ 
the total rate of two-photon emission obeys the scaling law $\propto \gamma^4$
\cite{pair beam}.
The two-photon emission of E1$\times$E1 type is usually weaker
than M1 type emission, although the M1 rate is much smaller than
the E1 rate.

\vspace{0.5cm} 
\section
{\bf Appendix B: Mathematical supplements}

\subsection
{\bf Bessel function of large orders for large arguments}

In order to calculate the phase integral of a type as
in eq.(\ref{phase integral}), we note the  integral representation of 
Bessel function that holds for non-integer $\nu$'s \cite{nist}:
\begin{eqnarray}
&&
J_{\nu}(z) = \frac{1}{\pi} \int_0^{\pi} d\theta \cos (\nu \theta - z \sin \theta)
- \frac{\sin \nu \pi}{\pi} \int_0^{\infty} dt e^{-\nu t - z \sinh t}
\,,
\end{eqnarray}
which holds for $\Re \nu > 0, \Re z > 0$.
The second integral is limited from above:
for real and positive $\nu, z$  by
\begin{eqnarray}
&&
| \int_0^{\infty} dt e^{-\nu t - z \sinh t}|
< \int_0^{\infty} dt e^{- ( \nu + z) t} = \frac{1}{ \nu + z}
\,,
\end{eqnarray}
which means that this second contribution is  small contribution in the large $\nu, z$ limit.
We then work out the other half of integration range,
\begin{eqnarray}
&&
\int_{\pi}^{2\pi} d\theta \cos (\nu \theta - z \sin \theta)
= \Re \left( \int_0^{\pi} d\theta' e^{i\nu (\theta' -\pi) + z \sin \theta')}
\right) 
\sim \Re e^{-i\nu \pi}J_{\nu}(-z) = J_{\nu}(z) 
\,.
\end{eqnarray}
Thus, by neglecting the sub-leading terms,
\begin{eqnarray}
&&
J_{\nu}(z)  = \frac{1}{2\pi} \int_0^{2\pi} d\theta \cos (\nu \theta - z \sin \theta)
+ O(\frac{1}{\nu + z})
\,.
\end{eqnarray}
The asymptotic behavior of the Bessel function,
\begin{eqnarray}
&&
J_{\nu}(\nu \sec \beta) = 
\sqrt{\frac{2 }{\nu \pi \tan \beta }} \cos (\nu \tan \beta - \nu \beta - \frac{\pi}{4}) 
\,,
\end{eqnarray}
can be used to derive for $0 < a^2 - b^2 \ll |b| \rightarrow \infty$
\begin{eqnarray}
&&
J_b(a) \sim \frac{1}{\sqrt{\pi}} (a^2 - b^2)^{-1/4}
\,.
\end{eqnarray}
This limiting behavior is valid when $a^2 - b^2 \gg 1$.
If  $a^2 - b^2 =O(1)$, a function $f(1/(a^2-b^2)\,)$ which may
be expressed in power series expansions multiplies.
Since $a^2- b^2 = O(\,(\rho \epsilon_{eg})^2)$,
the asymptotic form is usually excellent except at
special points of photon energies and emission angles.

In the following subsection we point out relevance of
the Bessel function to the Floquet system governed
by a linear set of ordinary differential equations
having periodic coefficient function.

\subsection
{\bf Appendix B: Floquet system}

We point out relevance of the problem to the Floquet system \cite{floquet}.
The function $Y(u) = \cos \Phi(u)$ along with $Z(u) = \sin \Phi(u)$ satisfies
\begin{eqnarray}
&&
\frac{d}{du}
\left(
\begin{array}{c}
Y  \\
Z   
\end{array}
\right)
= \Phi'(u) \left(
\begin{array}{cc}
0 & -1  \\
1 &  0
\end{array}
\right)
\left(
\begin{array}{c}
Y  \\
Z  
\end{array}
\right)
\,,
\\ &&
 \Phi'(u)  = 
b - a \cos (u - \theta')
\,.
\end{eqnarray}
Unlike $\Phi(u)$, its derivative $\Phi'(u)$ here is periodic.
The general theorem \cite{floquet} states that solutions $(Y, Z)$ are written by
\begin{eqnarray}
&&
\cos (\mu t) \;{\rm or}\; \sin (\mu t) \times P(t)
\,, \hspace{0.5cm}
P(u +T) = P(u)\;
( {\rm periodic \; function\; of\; period}\; T = \frac{2\pi}{\beta }\rho)
\,.
\end{eqnarray}

Eigenvalues $\lambda_j = i \mu_j \equiv \lambda \,, j=1,2$ are determined by solving
differential equations in one period under the boundary conditions
$(Y(0), Z(0)\,) = (1,0), (0,1)$.
In terms of solutions written in the matrix form $\Psi(t)$
the eigenvalue equation is
\begin{eqnarray}
&&
{\rm det}\; 
\left( \Psi(T) - \lambda
\right) = 0
\,.
\end{eqnarray}

\begin{figure*}[htbp]
 \begin{center}
 \epsfxsize=0.6\textwidth
 \centerline{\epsfbox{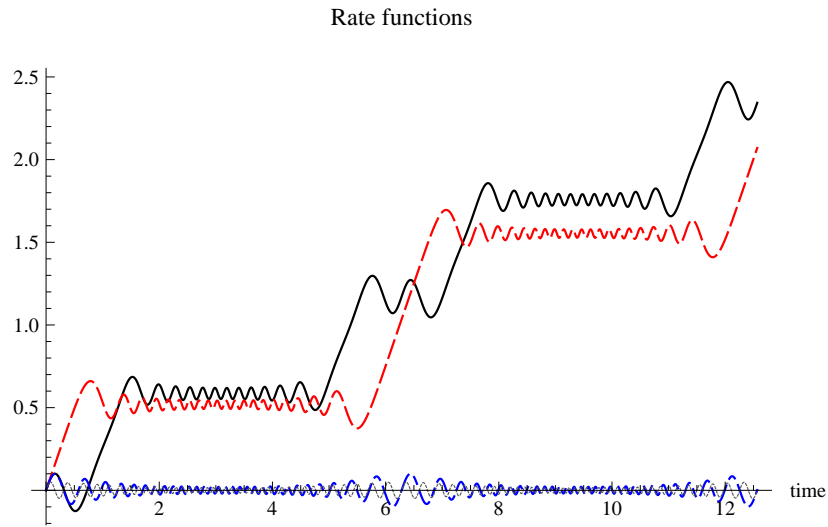}} \hspace*{\fill}
   \caption{ Rates in various $(b,a)$:
$(20\times 0.5, 20)$ in solid black,
(20, 20) in dashed red,
$ (20\times 1.5, 20)$ in dash-dotted blue,
and 
$(20\times 2, 20)$ in dotted black.
}
\label {compared rates}
 \end{center} 
\end{figure*}

\begin{figure}[htbp]
\begin{center}
\begin{minipage}{7.5cm}
     \epsfxsize=1\textwidth
 \centerline{\epsfbox{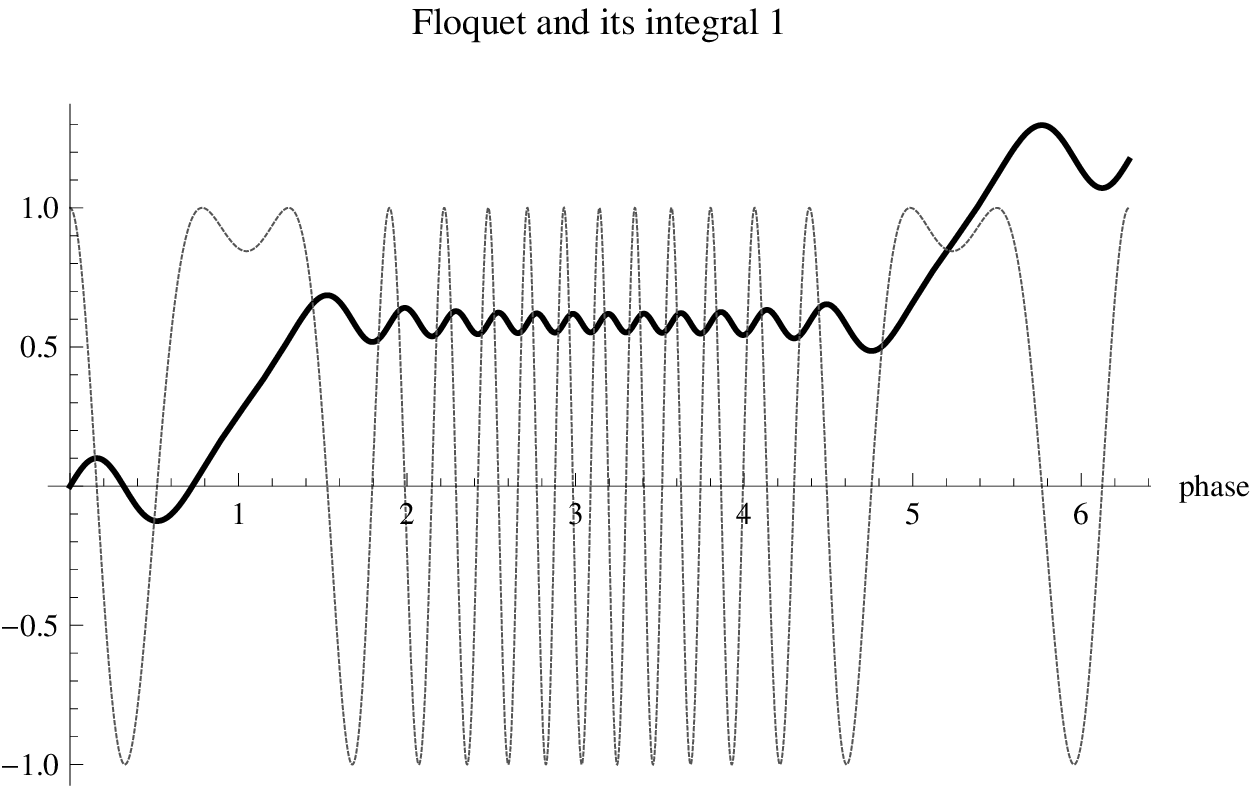}} \hspace*{\fill}
\end{minipage}
\begin{minipage}{0.5cm}
$\;$
\end{minipage}
\begin{minipage}{7.5cm}
   \epsfxsize=1\textwidth
 \centerline{\epsfbox{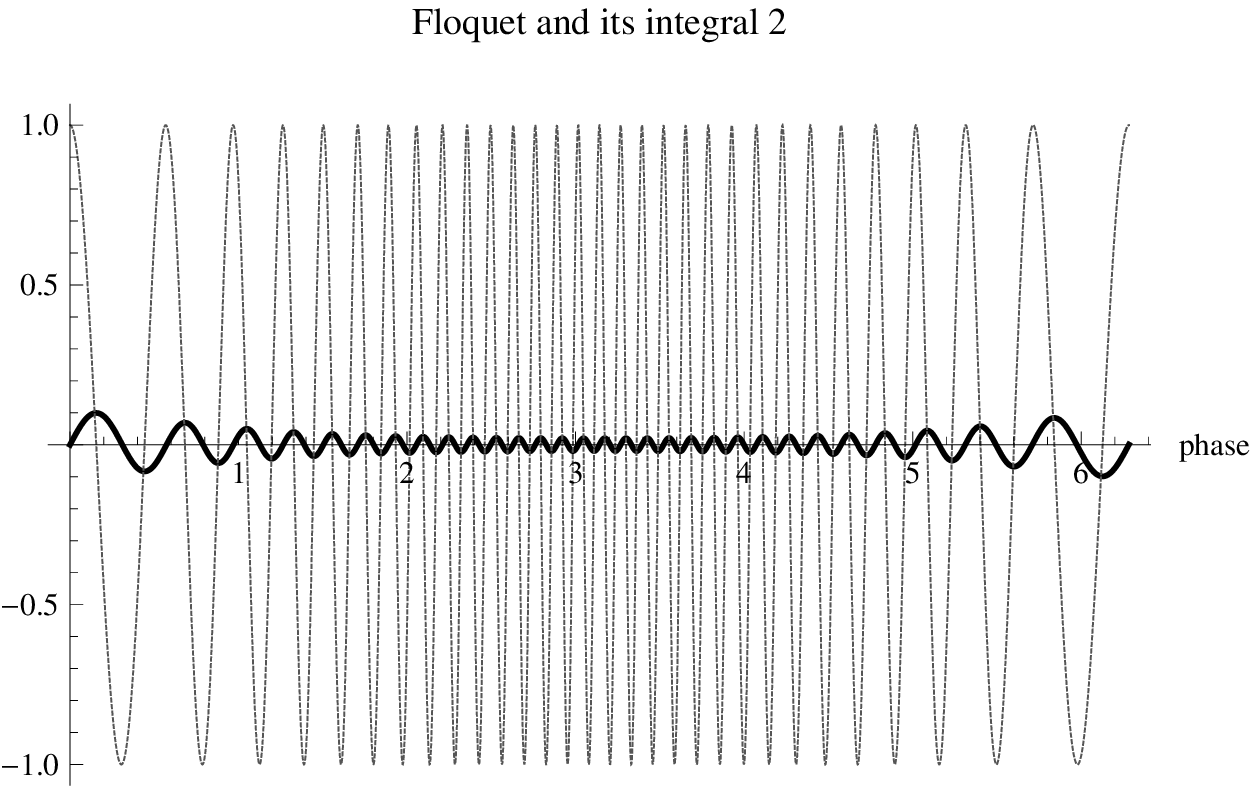}} \hspace*{\fill}
\end{minipage}
\caption{$\cos \Phi(u)$  and its integral for
a combination of  $(b,a)$
In the left pannel $(20\times 0.5, 20)$:
$\cos \Phi(u)$  in dotted black, and its integral  in solid black.
In the right panel $(20\times 0.5, 20)$.
}
\label {phase function and its integral}
\end{center}
\end{figure}

It is found that for large values of $b \sim a $ a large time integral 
may be obtained, but it depends on how close these two values  are.
It is important to differentiate two cases of $b \leq a$
and $b > a$ in which stationary points of $\Phi' = 0$ do or do not
exist.
Some cases of $b>a$ are illustrated in 
Fig(\ref{compared rates}).
To understand deeper,
we plotted the phase function $\cos \Phi(u)$ and its
integral $\int_0^u dy \cos \Phi(y)$
in Fig(\ref{phase function and its integral}),
one case showing a steady increase over one period
of circulation and the other case showing a failed
increase.
A nearly plateau-like region of the phase itself
at relatively early phase of circulation
is a crucial condition leading to the stable phase
of the rate integral.

Using the terminology of periodic potentials that one encounters
in solid state physics, one would say that
large rate integrals occur when
parameters $a,b$ (necessarily in the stability band due to
bounded functions of sinusoidal functions) are near the boundary to the
instability band.
The situation may be phrased in a different way.
One may introduce another function, namely the phase integral itself,
\begin{eqnarray}
&&
W (u)= \int_0^u dy \cos \Phi(y)
\,.
\end{eqnarray}
The three function system $(X,Y,W)$ now satisfies
a closed set of differential equations whose coefficient functions
are periodic.
This time the new W may not be bounded,
hence this Floquet system may belong to
the instability band for particular regions of $(a,b)$.
It would be useful to explore more of these features
from the point of  instability/stability band structure
and identify quantitatively the large rate region.

\vspace{0.5cm}
 {\bf Acknowledgements}

This research was partially supported by Grant-in-Aid for Scientific
Research on Innovative Areas "Extreme quantum world opened up by atoms"
(21104002) from the Ministry of Education, Culture, Sports, Science, 
and Technology, and JSPS KAKENHI Grant Number 15H02093.


\begin{thebibliography}{99}

\bibitem{schwinger}
J. Schwinger,
Phys. Rev.{\bf 75},1912(1949). 

J. Schwinger, Proc, Natl. Acad. Sci. {\bf 40},
132(1954).

\bibitem{pair beam} 
M. Yoshimura and N. Sasao,
Physical Review {\bf D92}, 073015(2015)
and arXiv: 1505.07572v2(2015).




\bibitem{nist}
NIST Handbook of Mathematical Functions,
Cambridge University Press.


\bibitem{floquet}
H. Hochstadt,
{\it Differential Equations},
Dover Publication (New York), Chapter 5 (1963).


\bibitem{asaka et al}
T. Asaka, M. Tanaka and M. Yoshimura,
"Basic oscillation measurables in the neutrino pair beam", 
paper in preparation.

\end{thebibliography}
\end{document}